\documentclass[preprint,review,12pt]{elsarticle}

\usepackage{amssymb}

\usepackage{gensymb}
\usepackage{bm}
\usepackage{subfig}
\usepackage{float}
\usepackage{url}
\usepackage{comment}
\usepackage{xcolor}
\usepackage{textcomp}
\usepackage{graphics}
\usepackage{array}

\journal{Carbon}

\begin{document}

\begin{frontmatter}



\title{Real time observation of glass-like carbon formation from SU-8 using X-ray and ultraviolet  photoelectron spectroscopy}


\author[inst2]{Simon Astley}
\author[inst1]{Jaspa Stritt}
\author[inst1]{Soumen Mandal}
\ead{mandals2@cardiff.ac.uk, soumen.mandal@gmail.com}
\author[inst1]{Jerome A. Cuenca}
\author[inst2]{D. Andrew Evans}
\author[inst1]{Oliver A. Williams}
\ead{williamso@cardiff.ac.uk}

\affiliation[inst2]{organization={Department of Physics},
            addressline={Aberystwyth University}, 
            city={Aberystwyth},
            country={UK}}
\affiliation[inst1]{organization={School of Physics and Astronomy},
            addressline={Cardiff University}, 
            city={Cardiff},
            country={UK}}
            
\begin{abstract}
The structural development and change in elemental composition of SU-8 3005 photoresist into glass-like carbon due to pyrolysis up to 1000~$\degree$C is investigated utilising \textit{in-situ} x-ray and ultraviolet photoelectron spectroscopy (XPS/UPS) under ultra-high vacuum (UHV). XPS spectra were analysed in order to investigate changes to elemental composition and physical structure. Peak asymmetry in the measured C 1s spectra is found to be a clear indicator of a transition in both physical structure and increased electrical conductivity. The \textit{in-situ} XPS measurement of pyrolysis is  effective in isolating changes in oxygen composition solely due to the pyrolysis process. Oxygen concentration, C 1s peak asymmetry and C 1s peak positions are strong indicators of semiconducting SU-8 transitioning to conducting glass-like carbon. For SU-8 pyrolysed above temperatures of 500~$\degree$C, a clear development is observed in the material structure and composition towards a carbon rich conducting network indicative of glass-like carbon. UPS spectra were analysed to investigate the changes in secondary electron cut-off (SECO) and valence band maximum (VBM) as the SU-8 layer is heated in UHV. The changes in SECO and VBM correlates well with the XPS data and a zero binding energy state is observed at 1000~$\degree$C.

\end{abstract}


\begin{highlights}
\item \emph{In-situ} observation of glass-like carbon formation.
\item X-ray photoelectron spectroscopy showing transformation of non - conductive SU-8 to a conductive thin film.
\item Observation of structural oxygen reduction between 300 and 600 $^o$C leading to the formation of a carbon rich network. 
\item Appearance of various features in UPS data in-line with compositional changes observed in XPS data.
\item Appearance of zero binding energy state at 1000~$\degree$C.
\end{highlights}


\end{frontmatter}



\section{Introduction}
\label{sec:intro}

\noindent
Glass-like carbon (GC) is a non-graphitising form of carbon having glass-like surface and exceptional properties including high-thermal stability, chemical resistance and electrical conductivity\cite{usk2021, vie2022}. It finds applications in electrochemistry\cite{pin2019, muru2022}, biosensors\cite{zhang1996,lak2022}, micro-electrical-mechanical systems (MEMS)\cite{jang_su-8_2022, martinez-duarte_su-8_2014, singh_pyrolysis_2002} etc due to its corrosion resistance\cite{usk2021, vie2022} and biocompatibility\cite{usk2021, vie2022}. Polymeric materials like SU-8 form the starting material for formation of GC through high temperature pyrolysis in an inert atmosphere.

\begin{figure}[H] 
    \centering
    \includegraphics[scale=0.5]{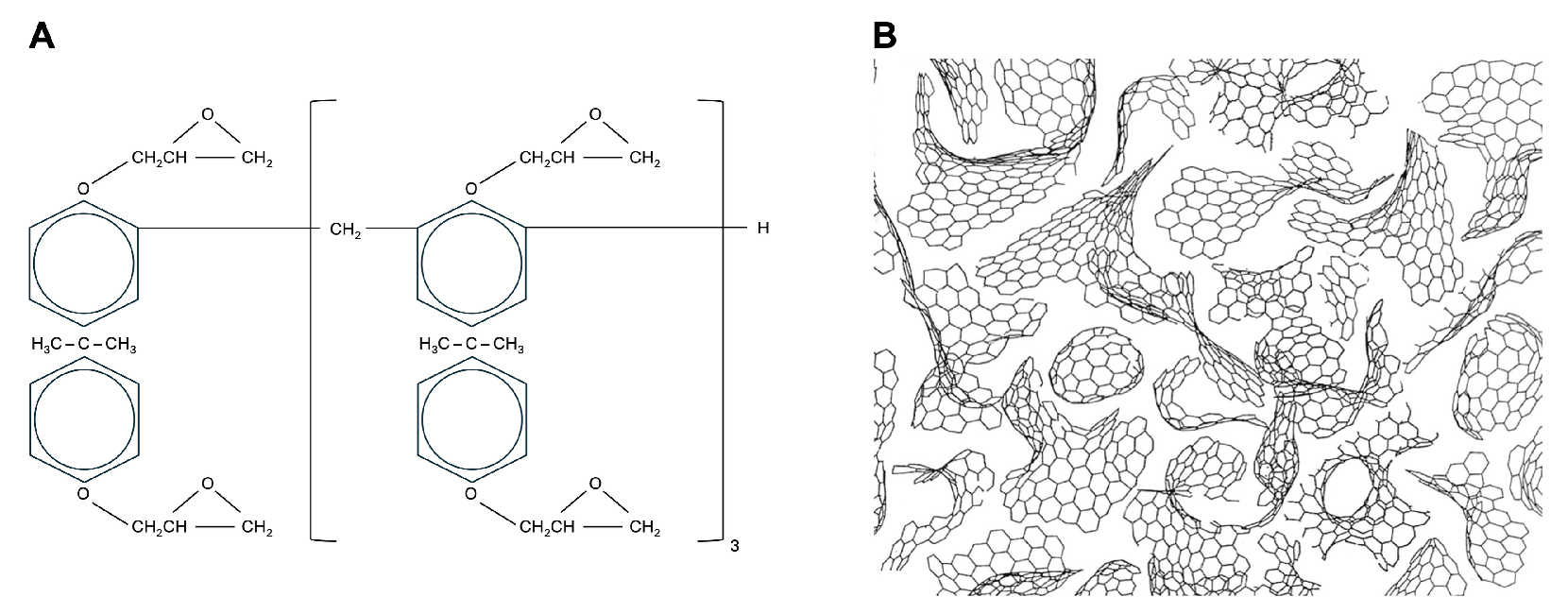}
    \caption{\textit{A) The chemical structure SU-8 molecule is shown here. (Adapted with permission from Genolet et al. \cite{geno99}) B) Schematic of glass-like carbon based on Harris' model \cite{harri13} where the the non-graphitising carbons are shown as an amalgamation of graphene and fullerene like structures. (Reprinted with permission from Harris\cite{harri13}). }}
    \label{gcs}

\end{figure}

The structure of glass-like carbon is typically characterised by either a ribbon-like arrangement \cite{jenkins_structure_1971, ban_lattice-resolution_1975} or a fullerene-related framework \cite{harris__fullerene-related_2004}. Both models describe its formation through the pyrolysis of an organic polymer, a process that preserves the morphology of the precursor while bypassing a plastic phase. The major steps of pyrolysis include i) Pre-carbonisation (elimination of solvent and unreacted monomer at temperatures below 300~$\degree$C), ii) Carbonisation (the elimination of oxygen and hydrogen as well as the development of an interconnected polymer-like sp$^2$ carbon structure at temperatures of 300~$\degree$C$<$T$<$1200~$\degree$C) and iii) Annealing (eliminating defects/impurities and further increases the inter-connectivity of the carbon network) \cite{martinez-duarte_su-8_2014}. Figure \ref{gcs} shows the structure of SU-8 molecule\cite{geno99} and a schematic of non-graphitising carbon based on Harris' model\cite{harri13}. The SU-8 is a negative tone photoresist consisting of eight equivalent epoxy and eight equivalent ether sites as shown in Figure \ref{gcs}A. Once pyrolysed it forms glass-like carbon which can be schematically represented as a amalgamation of graphene and fullerene like structures as described by the Harris' model\cite{harri13} (Figure \ref{gcs}B) and also reported by Sharma et al. \cite{sharma2018} using \emph{in-situ} transmission electron microscopy.

In this work X-ray photoelectron spectroscopy (XPS) and ultraviolet photoelectron spectroscopy (UPS) was used to study the real time carbonisation and annealing step in the formation of glass-like carbon from SU-8 photoresist in ultra high vacuum (UHV). 

\section{Experimental Methods}
\label{sec:methods}
SU-8 was originally developed by IBM \cite{shaw_negative_1997, lorenz_su-8_1997} as a photoresist material, sensitive to near-UV radiation (365nm). The photoresist cross links to produce a polymer structure through a process called cationic polymerisation, whereby the introduction of a positive ion results in the opening of an epoxy ring in order to allow for chemical bonding to epoxy groups of other SU-8 molecules \cite{olziersky_insight_2010}. SU-8 is typically dissolved in an organic solvent along with a photoacid generator (Triarylsulfonium/hexafluoroantimonate salt) that produces a strong acid (H+ ions) when exposed to UV-radiation \cite{olziersky_insight_2010}. Exposure to near-UV radiation produces H+ ions across the entire sample \cite{zhang_characterization_2001}. A post-exposure baking step is then used to promote a series of cross-linking reactions between epoxide groups in the exposed areas. The resultant material is a series of SU-8 molecules, each cross-linked to a maximum of 8 other molecules, resulting in a highly connected polymer network \cite{olziersky_insight_2010, lima_sacrificial_2015, sikanen_characterization_2005}.


\subsection{Sample Preparation}

 In this study, SU-8 3005 photoresist was spun onto 10 x 10~mm silicon substrates at a speed of 2000~rpm for 45 seconds followed by a soft bake at 110~$\degree$C. The samples were then flood exposed to broadband UV light from a mercury-xenon gas lamp in order to form cross links between the epoxy groups to form a large polymer network followed by a post-exposure bake also at 110~$\degree$C. The samples were then heat treated to 300~$\degree$C in order to remove any solvent and photosensitiser. The samples were placed into an Inconel Alloy 600 tube furnace. The tube then underwent 3 purge cycles of being evacuated to 2 mbar and flooded with argon up to 500~mbar. After purging, the furnace was once again evacuated to 2~mbar before introducing a constant flow of 200~SCCM of argon and maintaining a pressure of 50~mbar. The samples were heated from room temperature to 300~$\degree$C at a rate of 2~$\degree$C/min and held at this temperature for 1~hour. The samples were then allowed to cool back to room temperature before being removed from the tube furnace.

\subsection{X-ray photoelectron spectroscopy}
\noindent 
C 1s and O 1s core level spectra were measured using XPS in real-time \cite{langstaff_system_2010} to determine changes in chemical composition of SU-8 through pyrolysis via the desorption of oxygen, restructuring of cross-linked SU-8 into a conducting sp$^2$ state, and conductivity of the sample through the asymmetry of the C 1s peak. Photoelectrons were collected and energy analysed  using a SPECS Phoibos 100 analyser with a PreVac twin anode X-Ray flood source providing Mg K$\alpha$ un-monochromated x-Rays at a 54$\degree$ angle, at a base pressure of 6~$\times$~10$^{-10}$ mbar. The data were collected using a 3~$\times$~7 mm slit and a 5 mm IRIS to ensure the information collected was purely from the SU-8 sample. The sample was measured as loaded and then during each annealing cycle. The sample was not sputtered due to potential damage to the SU-8, which would affect the potential onset of the conductivity. The sample was spot-welded using tantalum strips on the edges to a tantalum sample plate to ensure good electrical contact and to minimise charging of the sample.   

The sample was heated in 100~$\degree$C/50~$\degree$C steps from room temperature to 1000~$\degree$C using a graphite-boron nitride heater with the temperature measured by a thermocouple attached to the sample holder (calibrated using in-situ Raman spectroscopy). For each temperature, the core levels were measured at temperature at a pass energy (E$_p$) of 20~eV, with the survey scans measured at 100~eV. C 1s spectra were measured for the full temperature range, whilst O 1s spectra were measured until a peak was no longer observable. The spectra were then fitted with a Voigt function \cite{schm2014}, with the Lorentzian term restricted based on known line widths previously measured on the system. The 1000~$\degree$C measurements were taken whilst the sample was cool due to high pressures in the chamber, and in order to achieve better energy resolution for the final measurement.

\subsection{Ultraviolet photoelectron spectroscopy}
\noindent
UPS was conducted using a SPECS UVS3000 using a He I emission, and the sample was biased at -5 V to reliably measure the secondary electron cut-off (SECO). The sample was annealed up to 1000~$\degree$C with the same temperature steps as the XPS in the same system. Spectra were recorded at each temperature, except for 1000~$\degree$C where spectra were obtained at both 700~$\degree$C and room temperature, due to the higher pressures in the analysis chamber at 1000~$\degree$C. 
\section{Results \& Discussion}
\label{sec:results}
\begin{figure}[H] 
    \centering
    \includegraphics[scale=0.25]{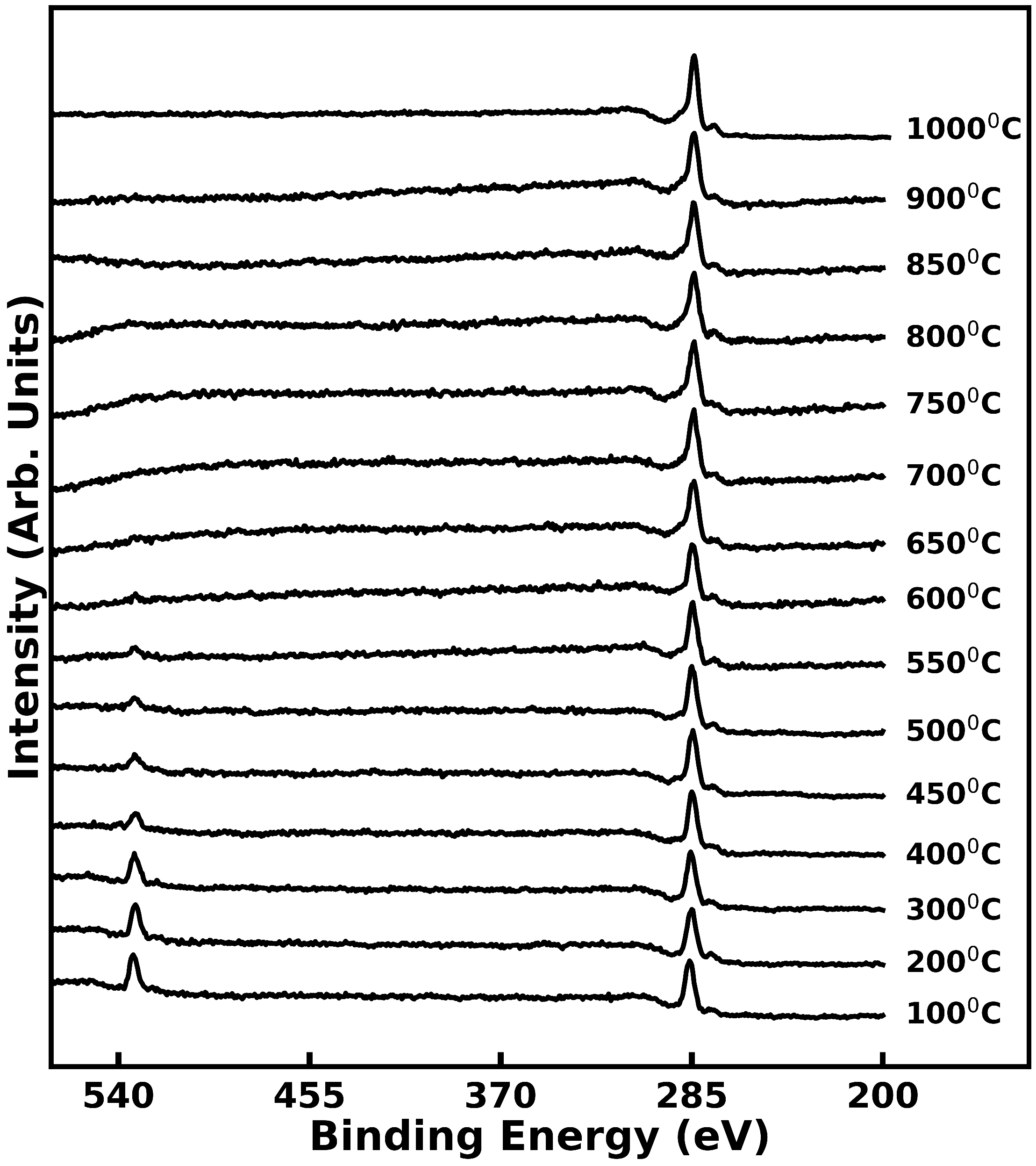}
    \caption{\textit{XPS Survey Scans for SU-8 3005 photo-resist pyrolysed at varying temperatures in UHV}}
    \label{fig:surveys_compiled}
\end{figure}

\begin{figure}[H] 
    \centering
    \includegraphics[scale=0.2]{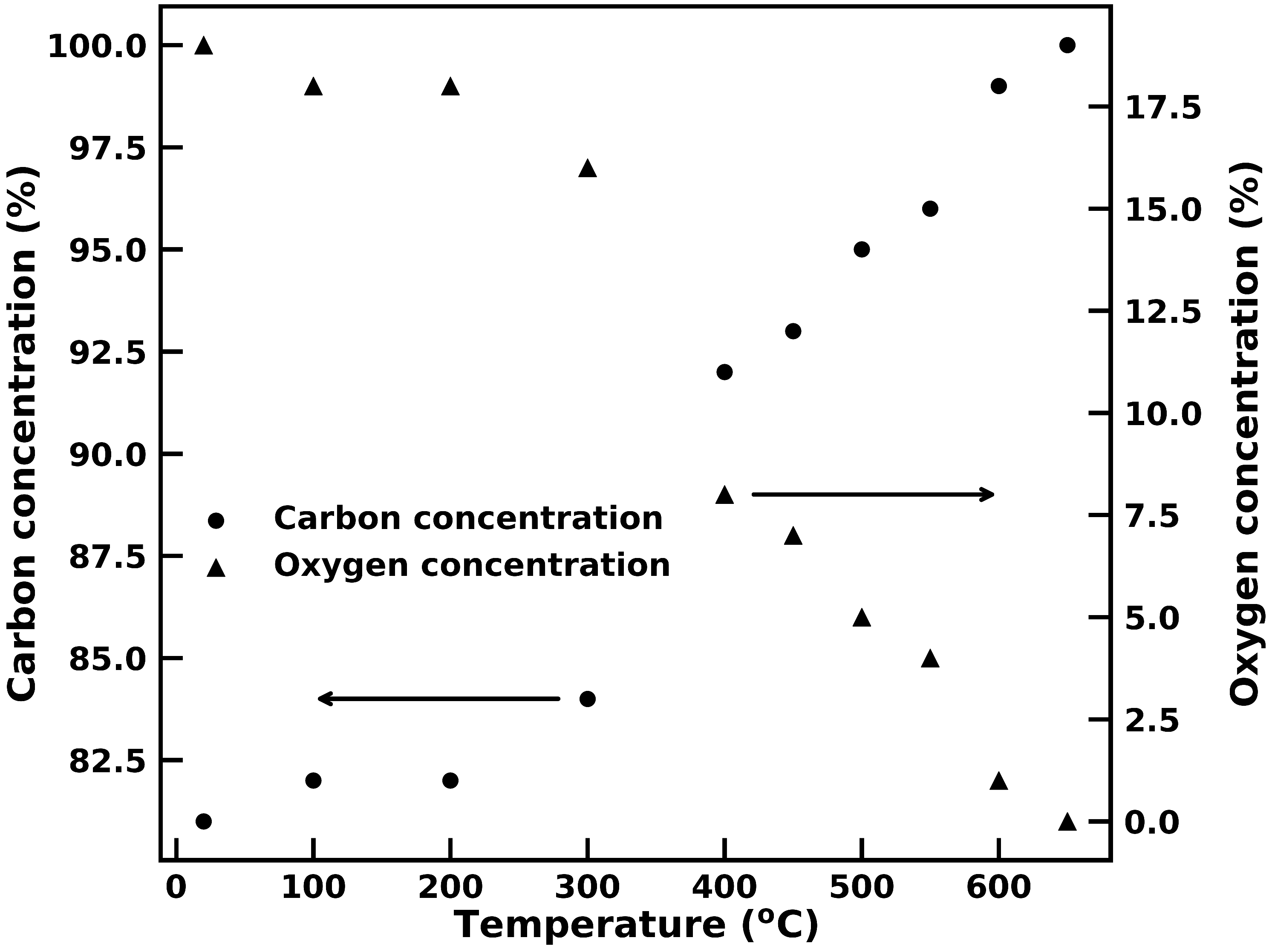}
    \caption{\textit{Elemental composition for carbon and oxygen of pyrolysed SU-8 3005 photoresist as measured from C 1s and O 1s core level peak intensities.}}
    \label{figo1s}
\end{figure}

\noindent
Figure \ref{fig:surveys_compiled} shows the compiled XPS survey scans for SU-8 pyrolysed in UHV at temperatures between 100 - 1000~$\degree$C. C 1s core level emission peaks at $\sim$285~$eV$ are present throughout the entire pyrolysis process, with the intensity increasing with pyrolysis temperature (T$_P$). O 1s features can be seen at $\sim$533~$eV$ at temperatures below 650~$\degree$C, steadily decreasing in intensity as T$_P$ increases and the oxygen is desorbed. The elemental composition, calculated from the relative C 1s and O 1s peak intensities using CASAXPS software \cite{fairley2021}, is shown in figure \ref{figo1s}. 

\begin{figure}[H] 
    \centering
    {{\includegraphics[scale=0.3]{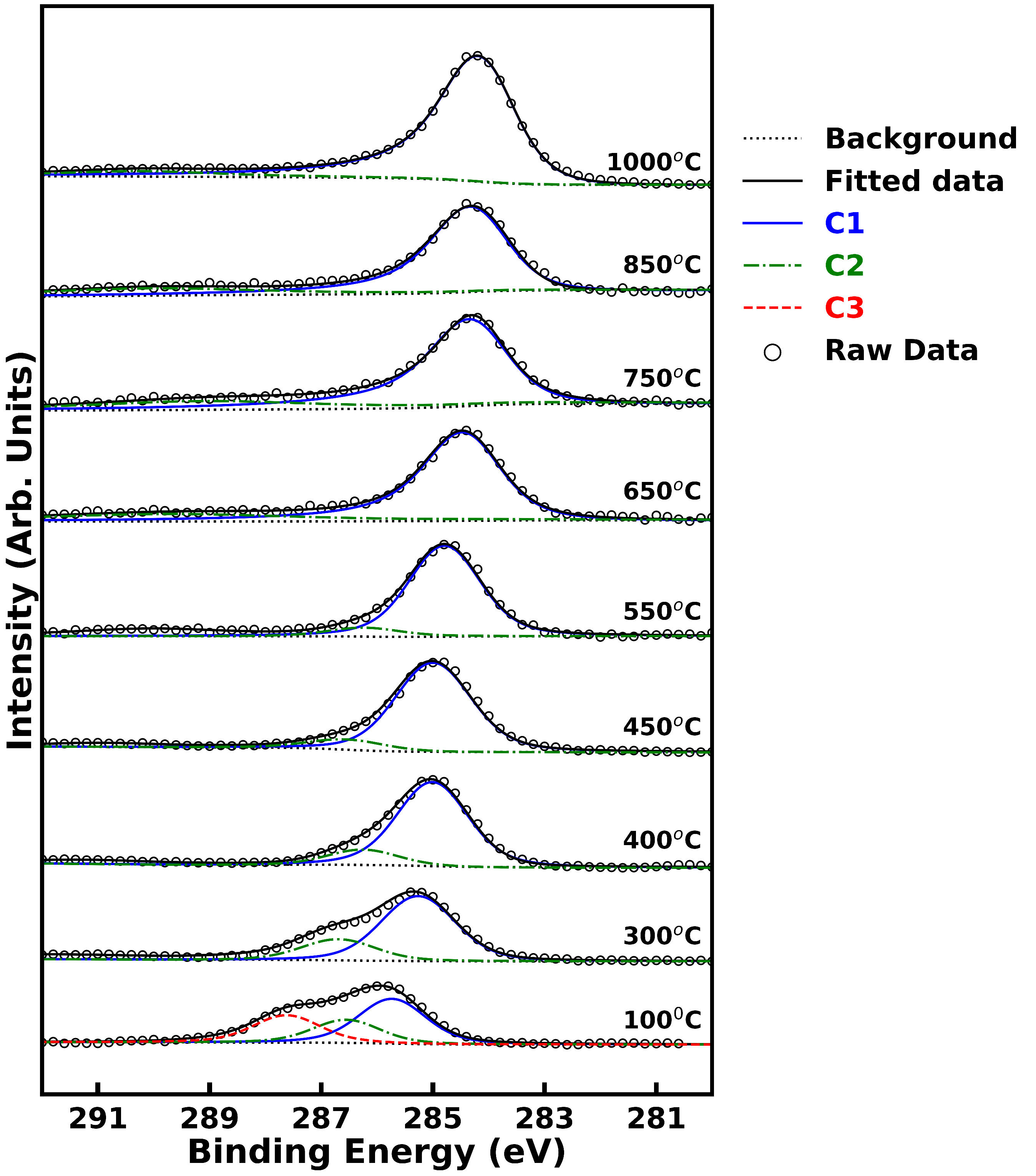} }}
    \caption{\textit{C 1s core level emission spectra as a function of increasing temperature in-vacuo. The spectra have been fitted using multiple Voigt curves relating to various bonding groups within the material. The C1 represents C-C and C-H groups, C2 represents C-O groups and C3 represents contaminant carbon.}}    
    \label{c1s}
\end{figure}

\begin{figure}[H] 
    \centering
    
    {{\includegraphics[scale=0.3]{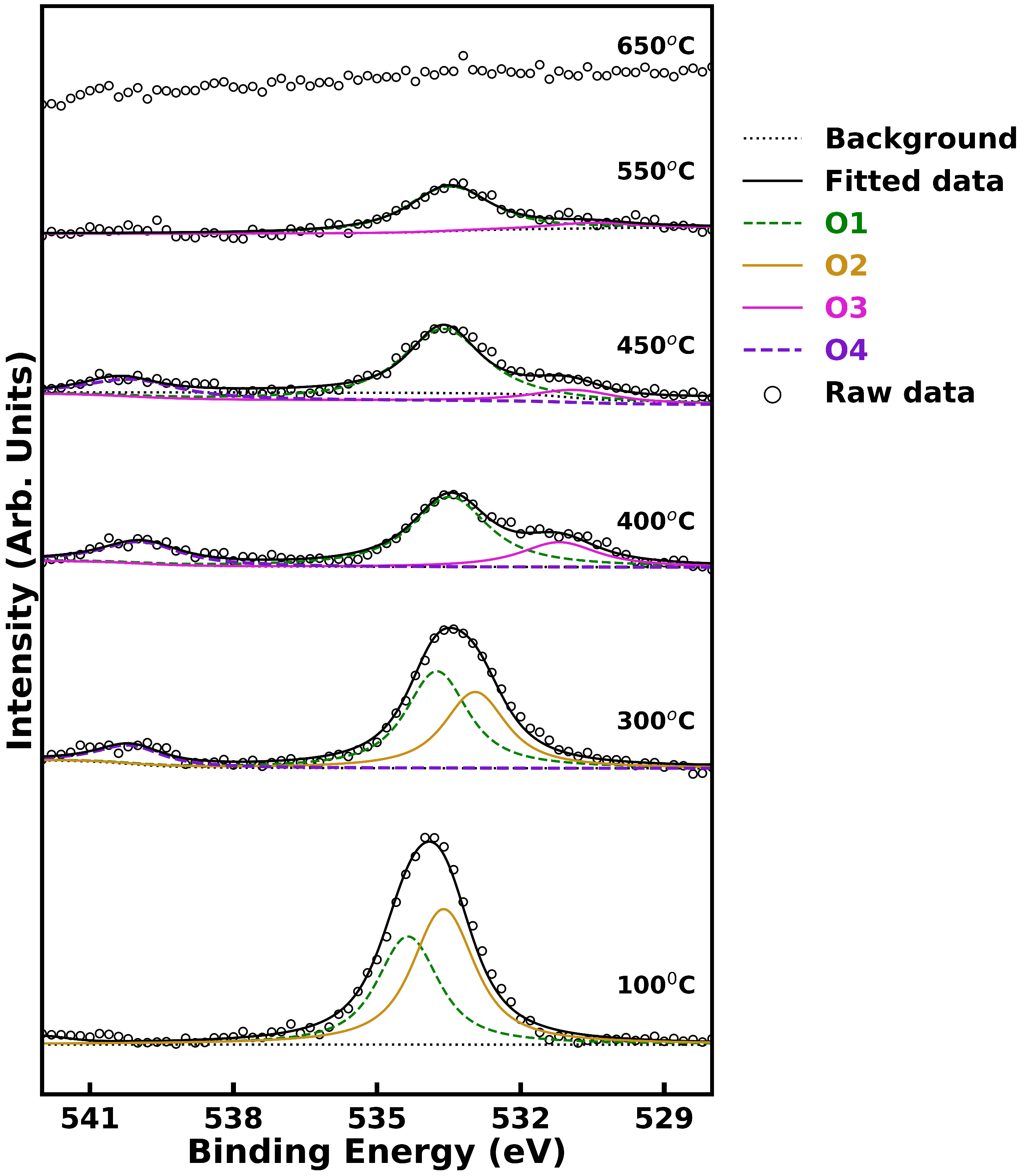} }}
    \caption{\textit{O 1s core level emission spectra as a function of increasing temperature in-vacuo. The spectra have been fitted using multiple Voigt components relating to various bonding groups within the material.O1 and O2 represent the two oxygen environments in the SU-8 molecule, O3 represents C=O sites that are formed at 400 C, and O4 is not associated with the SU-8 molecule. }}    
    \label{o1s}
\end{figure}

\noindent
The structural development of glass-like carbon, from SU-8 through pyrolysis, arises from the elimination of non-carbon atoms. The SU-8 molecule contains 87 carbon atoms and 16 oxygen atoms, resulting in an expected oxygen concentration of 15.5\%. As shown in Figure \ref{figo1s}, below 300~$\degree$C the oxygen concentration is shown to be higher than this (reducing from $\sim$19\% to $\sim$16\% between 20 - 300~$\degree$C), suggesting the existence of oxygen contamination from other sources. At 300~$\degree$C therefore, XPS intensity analysis suggests that the SU-8 is contamination free. At lower temperatures, the amount of contaminants is process-dependent and the surface is not reproducible until heated to 300~$\degree$C in UHV. Previous studies have shown that the pyrolysis step of pre-carbonisation occurs at temperatures below 300~$\degree$C and involves the elimination of solvents and unreacted monomers \cite{martinez-duarte_su-8_2014}, with carbonisation only beginning at 300~$\degree$C. The oxygen concentration then decreases rapidly between 300 - 600~$\degree$C, with no measurable O 1s signal present after 600~$\degree$C. This is again in-line with previous pyrolysis studies showing that the majority of oxygen elimination occurs in the earlier stages of carbonisation (300 - 500~$\degree$C) \cite{lee_mechanisms_1964}. Figure \ref{figo1s} illustrates the overall oxygen desorption to be continuous between 300~$\degree$C and 600~$\degree$C.
\\\\
Figures \ref{c1s} and \ref{o1s} shows detailed XPS spectra at binding energies in the C 1s (280 - 292~eV) and O1s (528 - 542~eV) regions respectively. Selected C 1s and O 1s spectra are shown which have been fitted using Voigt functions (a convolution of a Gaussian and Lorentzian line-shape). A minimum number of peaks were selected for each spectrum, using physically meaningful variables for Gaussian and Lorentzian contributions related to the sample and  instrumentation used. The fitting process was iterated to minimise residuals. An asymmetry parameter was introduced to reflect the changing conductivity from semiconducting to conducting and therefore, flexibility in the asymmetry fitting was allowed for the C 1s feature. The relative Gaussian and Lorentzian ratios were kept consistent throughout whilst allowing for broadening due to measurements at high temperatures. 
\\\\
The C 1s core level emission spectra (Figure \ref{c1s})  were fitted with a combination of peaks that are associated broadly with C-C, C-H and C-O bonding within the SU-8 molecule and, at higher binding energy, contaminant carbon atoms not associated with the SU-8 sample. Due to the instrumental energy resolution and the semiconducting nature of the sample, it is not possible to ascribe precise bonding environments to the core level components \cite{chen_review_2020, brunetti_xps_2012}. The highest binding energy component (C3), present at 100~$\degree$C is identified due to contaminant carbon species that are not associated with the SU-8 molecule. These are desorbed as temperature increases, with no contributions  at temperatures above 300~$\degree$C. The core level line-shape and position is not reproducible for different samples as-loaded, following ex-situ preparation. However, the XPS spectra are consistent for all samples following annealing to 300~$\degree$C in UHV which suggests clean, contamination-free SU-8. Fitting of the lower-temperature surfaces is thus more subject to variation and less representative of the SU-8 molecule. The C 1s spectrum at 400~$\degree$C can be adequately fitted with two components as shown in Figure \ref{c1s}. The lower intensity component (C2) is around 10\% of the more intense component (C1) and this is similar to the overall oxygen concentration  at this temperature (Figure \ref{figo1s}).  The component at a binding energy of $\sim$286.7 eV (C2 in Figure \ref{c1s}) represents a number of possible C-O bonding environments in the matrix.  C2 is present at 300~$\degree$C and it is not present at the higher temperatures. This is interpreted as the desorption of oxygen from the SU-8 as the glass phase is formed. The main C 1s component at a binding energy of 285 eV is due to carbon atoms bound to other carbon atoms and to hydrogen atoms within the SU-8 molecule ( C1 in Figure \ref{c1s}). Within the energy resolution of the instrument, it is not possible to unambiguously identify the different C-C bonding sites nor C-H bonding where the carbon atoms have similar symmetry. The binding energy of these components are close together for other carbon containing materials \cite{yamada13, gohda20} and, in a combined experimental and theoretical study of ether and epoxy containing material, the binding energy for these two sites are reported to be the same \cite{fuji16}.  For semiconductors, core level binding energy can also be affected by local band bending and changes in band gap that can further contribute to the overall peak position and peak broadening \cite{astley_identifying_2022}. At lower temperatures, a further component at higher binding energy (C3) is required to obtain a reasonable fit. The relative intensity of this component is preparation-
dependent and, since it is not present at 300~$\degree$C, is ascribed to contaminant carbon atoms or organic fragments not associated with the SU-8 molecules (Figure \ref{c1s}). 
In table \ref{tab1} the FWHM of the C 1s main component has been summarised. The FWHM of the C 1s core level at the lower temperatures, corresponding to molecular SU-8, is broader than typical values measured in our XPS system for single phase carbon materials such as diamond and graphite (1.50eV vs. 0.80eV). This reflects the presence of multiple components of binding energies that are too close together to be resolved. As the temperatures increases, the FWHM is expected to increase due to thermal broadening. For SU-8, the FWHM increases initially with annealing, and at 700~$\degree$C, the FWHM of the main C 1s peak increases further due to different chemical environments, or non-sp$^2$ carbon, following the removal of oxygen groups. During cooling, the observed decrease in the FWHM is due to a reduction in thermal broadening, since the high temperature structural and chemical changes are preserved. This is evident in a comparison between the sample measured at 700~$\degree$C, before and after heating to 1000~$\degree$C. The reduced FWHM indicates a fully transitioned network. The C KLL Auger electron spectra yield a D parameter that is consistent with a structure that may contain multiple types of C-C bonds but has general ordering and high levels of sp$^2$ carbon. At the highest temperatures, the FWHM for the graphitic network is broader than the HOPG due to a combination of thermal broadening and multiple components that cannot be resolved into clear components. These could be due to defects and structural variation, such as non-hexagonal carbon\cite{barinew09,susinew2014,kimnew2015,diananew2021,kimnew2021,satonew2024}.

Since the identification of precise C-O bonding is not possible from the C 1s emission spectra even for O-terminated, single crystal diamond,  \cite{dontschuk_x-ray_2023} the O 1s core level spectra have been fitted as shown in Figure \ref{o1s}  for temperatures up to 650~$\degree$C. Each spectrum has been fitted with three peaks (O1, O2 and O3  in Figure \ref{figo1s}) with an additional peak, O4, not associated with the SU-8 molecule, at $\sim$540~eV.

\begin{center}
\begin{table}

\begin{tabular}{ | m{5cm} | m{5cm} | }
 \hline
 Temperature ($^o$C) & C1s main component FWHM (eV) \\ 
  \hline
 300 & 1.50 \\ 
  \hline
 500 & 1.55   \\  
  \hline
  700 & 1.79   \\  
  \hline
  700 post 1000 & 1.56   \\  
  \hline
  500 post 1000 & 1.53   \\  
  \hline
  300 post 1000 & 1.52   \\  
  \hline
  RT post 1000 & 1.49  \\  
  \hline
  HOPG (post 300) & 1.30   \\  
  \hline
\end{tabular}

 \caption{FWHM of the C1s main component at various temperatures.} \label{tab1}
\end{table}
\end{center}

\newpage

\noindent 
As shown in Figure \ref{gcs}A, there are two oxygen environments in the 
SU-8 molecule : one attached to an aromatic ring in a R-O-R ether configuration (O1) and the other in an epoxy configuration (O2) \cite{olziersky_insight_2010, lima_sacrificial_2015}. Each of these could have a distinct binding energy and thermal degradation mechanisms \cite{lee_mechanisms_1964}. At 300~$\degree$C, the fitted curve only requires the two components, O1 and O2, corresponding to the ether and epoxy sites in the SU-8 molecule. The binding energy difference is small due to the similarity in the oxygen environment where each oxygen atom is singly-bonded to two carbon atom. The intensity of the epoxy oxygen component is lower, suggesting that a preferential desorption has commenced at 300~$\degree$C. By 550~$\degree$C, only the ether oxygen component remains, and this disappears by 650~$\degree$C as these oxygen atoms are desorbed from the molecule. At 400~$\degree$C, a new component in the O 1s core level emission spectra at lower binding energy is required to obtain a reasonable fit (O3). Components at these binding energies are usually identified as due to double-bonded oxygen atoms in a ketone configuration \cite{giesbers_simulation_2013}. The binding energy of oxygen atoms in groups such as epoxy and ether can vary considerably for different molecules \cite{fuji16} and these authors also point out that charging can also lead to ambiguous assignment of sp$^3$ carbon. Comparison with other methods such as IR and Raman can confirm the presence of different oxygen species \cite{yamada13, gohda20, fuji16}. In the present study, the fitting parameters for the O 1s and C 1s core levels are consistent with the literature and the accepted model for SU-8 molecule discussed in Figure \ref{gcs}a. The peak located at $\sim$540.4~eV is not commonly observed for solids containing carbon and oxygen, but is close in energy to gaseous, molecular oxygen (O$_2$) observed in near-ambient
XPS \cite{astley_identifying_2022, axnanda_direct_2013, kim_adsorbate-driven_2018, avval_oxygen_2019, jakub_adsorbate-induced_2020, jeong_near_2021}. Oxygen is released during the transition from semiconducting to conducting phases and this oxygen desorbing from the structure at a slow rate may be detectable due to the porosity of the material. The O3 component is therefore  likely to arise  from the conversion of epoxy oxygen in the outer arm to ketone-like oxygen by the removal of a CH$_2$ group. At 550~$\degree$C, the ketone oxygen is desorbed and the aromatic rings start to fuse into the graphitic structure shown in Figure \ref{gcs}b. At temperatures below 300~$\degree$C there is likely to be a contribution to  the O 1s core level spectra from oxygen contaminants or fragments and not due to the SU-8 molecule. The 100~$\degree$C spectra in Figure \ref{figo1s} has been fitted using two components that are likely to contain contaminant and SU-8 oxygen sites.

\begin{figure}[H] 
    \centering
   {{\includegraphics[scale=0.4]{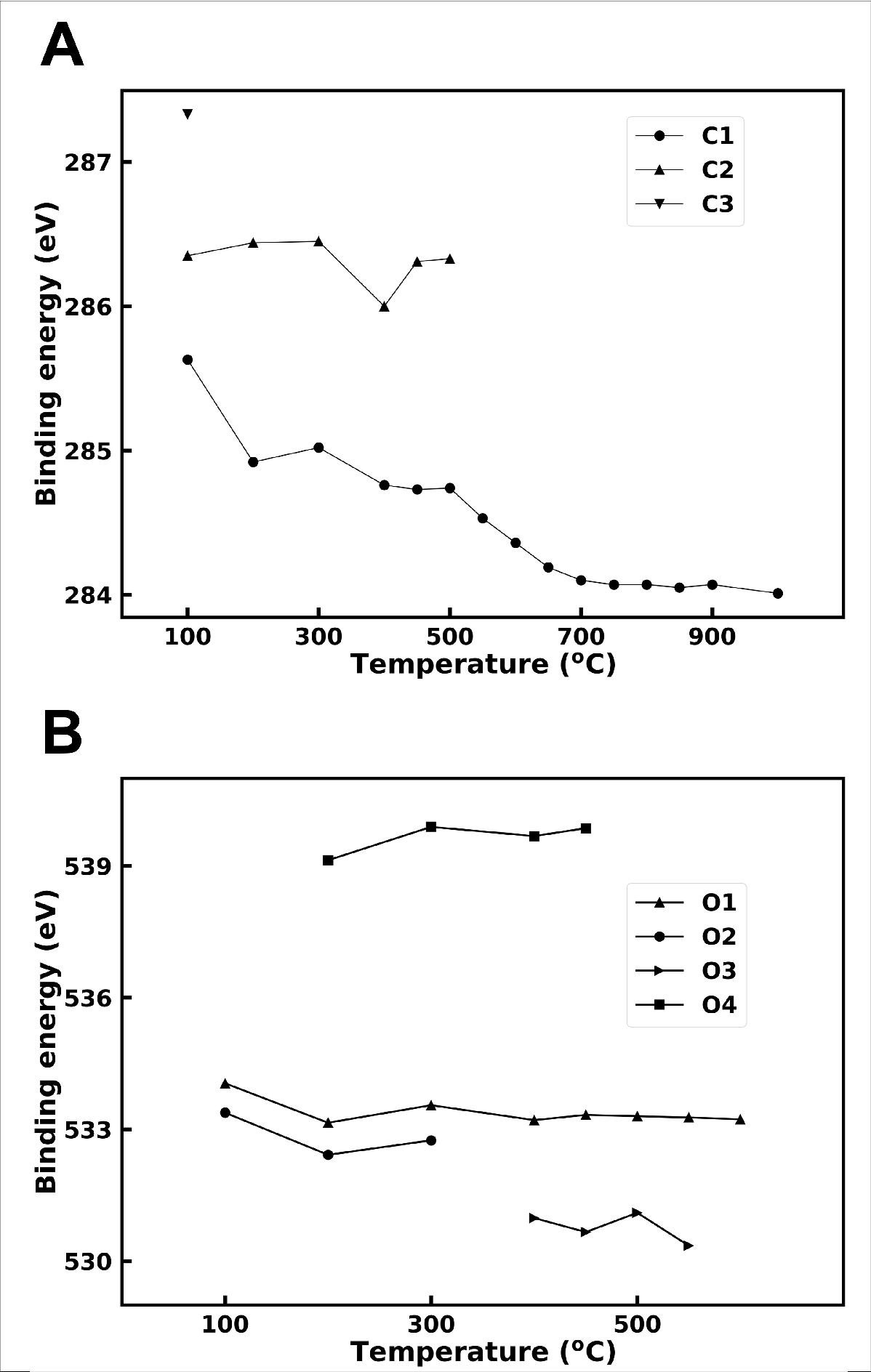} }}
    \caption{\textit{Peak positions plotted against pyrolysis temperature for C1s fitted components (C1-C3) (A) and for O 1s fitted components (O1-O4) (B)}}    
    \label{fig:PeakPositions}
\end{figure}

The temperature-dependence of the binding energy of each fitted peak for both the C 1s and O 1s regions is shown in Figure \ref{fig:PeakPositions}. In the C 1s spectra, contamination carbon is only present at low temperatures and is removed at temperatures above 100~$\degree$C. The position of the peak associated with C-O bonds in the C 1s spectra is largely unchanged, and this can be also seen in the O 1s spectra with all three O groups (Figure \ref{fig:PeakPositions}B), indicating little change in binding energy throughout pyrolysis. The main peak, dominated by C-C bonding,  in the C 1s spectra exhibits the most considerable change throughout the pyrolysis process. An initial decrease in binding energy between 100 - 300~$\degree$C is followed, at temperatures above 500~$\degree$C, by a further decrease in binding energy from 285.0 eV to 284.2 eV. Shifts in core levels can be due to a variety of physical and chemical changes \cite{astley_identifying_2022} whilst changes in the relative component intensities can be used to inform the processes and mechanisms for these shifts. The final peak position for the main C 1s component (C1) at a binding energy of 284.2 eV is indicative of a carbon structure similar to that of sp$^2$ carbon \cite{chen_review_2020, lascovich_evaluation_1991, morgan_comments_2021}. This suggests that whilst the elimination of oxygen during carbonisation has little effect on the structure, the elimination of hydrogen at temperature above 500~$\degree$C is key to development of an sp$^2$ network indicative of glass-like carbon.

\begin{figure}[H] 
    \centering
    \includegraphics[scale=0.25]{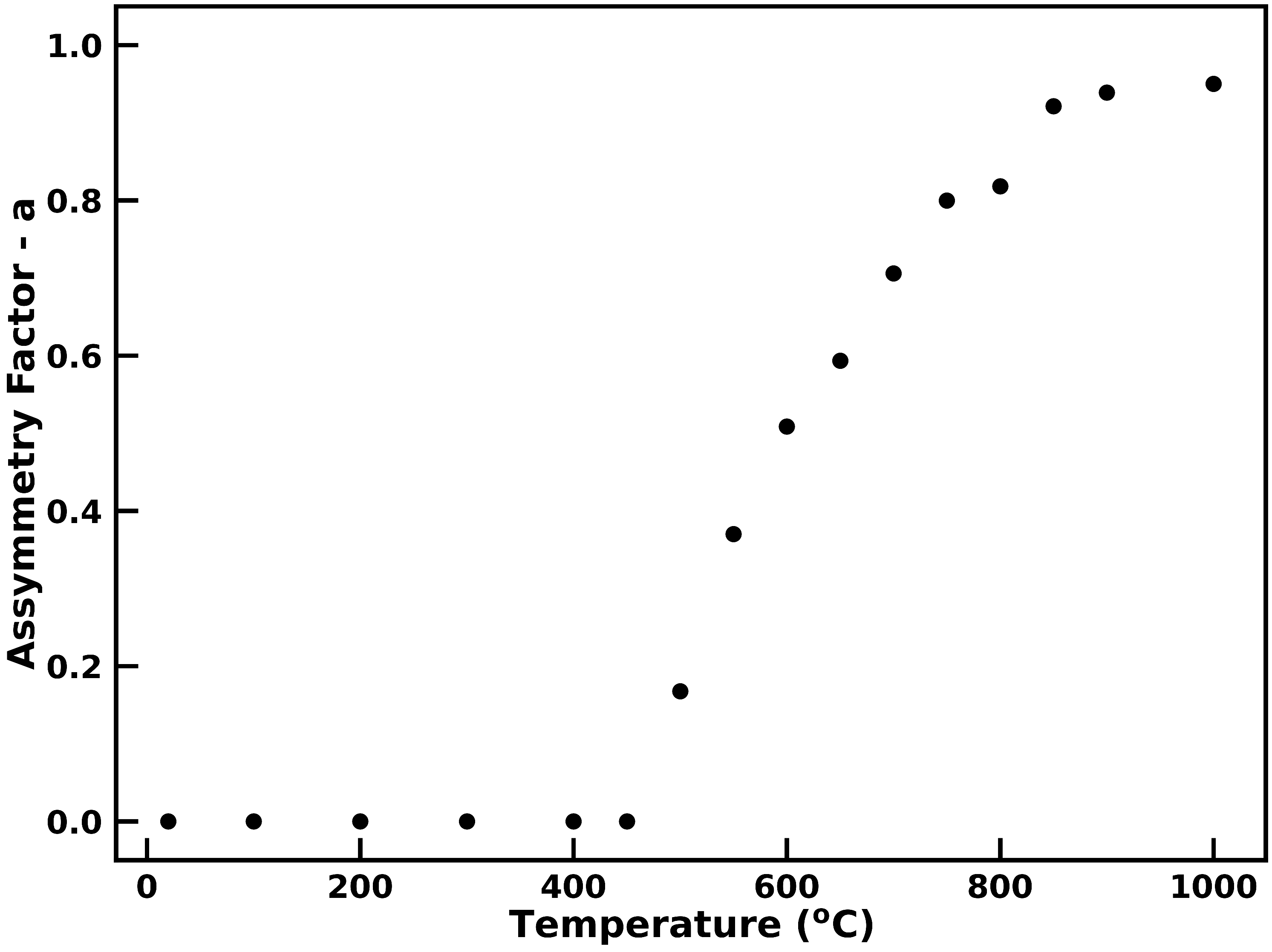}
    \caption{\textit{Asymmetry in the C1 peak in C 1s spectra, indicative of sp$^2$ carbon formation and an increase in electrical conductivity.}}
    \label{fig:asymmetry}
\end{figure}

The shift in binding energy thus indicates the transition from a semiconducting to conducting material between 500~$\degree$C and 700~$\degree$C. The intensity of the O 1s peak is low at these temperatures and hence it is not possible to determine with accuracy the binding energy of oxygen components in this temperature range. For all measurements of SU-8 at 300~$\degree$C and its conversion into a graphitic network at higher temperatures, are not affected by charging due to the photoemission process. Conductivity measurements \cite{stritt_development_2024} on this material yield a centre to edge resistance of 325k$\Omega$ at 700~$\degree$C and 4k$\Omega$ at 750~$\degree$C. From top to bottom the corresponding resistances are 0.46$\Omega$ to 0.03$\Omega$ between 700~$\degree$C and 750~$\degree$C respectively. For typical values of photocurrent generated  in the XPS spectrometer, significant charging shift would only be observed for a resistance of greater than 100M$\Omega$. The initial large shift in binding energy in the C 1s and O 1s, at temperatures up to 300~$\degree$C, suggests that the presence of contaminants affects the electronic nature of the measured sample surface changing the band bending or band gap of the semiconducting SU-8, confirming that a true measurement of SU-8 is not achieved until 300~$\degree$C.      
\noindent 
The C1 peak exhibits an increase in asymmetry during in-situ heating to higher pyrolysis temperatures. Figure \ref{fig:asymmetry} shows the variation of the asymmetry factor ($a$, as defined by the Voigt fit) with pyrolysis temperature. A sharp increase in asymmetry occurs at  550~$\degree$C that continues to a value of 0.95 at a pyrolysis temperature of 1000~$\degree$C.

 Asymmetric peaks are due to a variety of factors such as vibrational modes, multi electron excitations and electron - hole pair generation. In carbon structures such as graphite, the asymmetry can be ascribed to electrons excited into the conduction band during the photoemission process, causing a loss of energy for the core level photoelectrons. As the sample becomes more conductive, the change in asymmetry can be correlated with changing conductivity \cite{van1979, che1982, moe22}. 
\noindent
For a conductive carbon material, such as sp$^2$ carbon, an asymmetry in the C 1s line shape is observed towards higher binding energies. In highly conductive materials, core holes are screened by the relaxation of electrons in the conduction band. This results in ejected photo electrons losing energy to this initial excitation which manifests as higher binding energy events \cite{blume_characterizing_2015, susi_x-ray_2015, gengenbach_practical_2021}.

\noindent
The increase in asymmetry starting at 500~$\degree$C clearly demonstrates the two distinct stages of carbonisation. In the first stage (between 300-500~$\degree$C), most oxygen atoms are eliminated from the structure, leaving behind a conjugated system of carbon, hydrogen atoms and some ether based oxygen atoms. The second stage (between 500-1200~$\degree$C) involves the elimination of all other non-carbon atoms (including hydrogen and any remaining oxygen) resulting in a network of sp$^2$ carbon. As more non-carbon elements are eliminated, the network becomes highly interconnected and a rapid increase in conductivity can be observed \cite{stritt_development_2024}, manifesting as an increase in asymmetry. Similar formation of various interconnected sp$^2$ carbon species were observed in an \emph{in-situ} transmission electron microscopy  study of the pyrolysis of SU-8 \cite{sharma2018}.

\begin{figure} 
    \centering
    \includegraphics[scale=0.4]{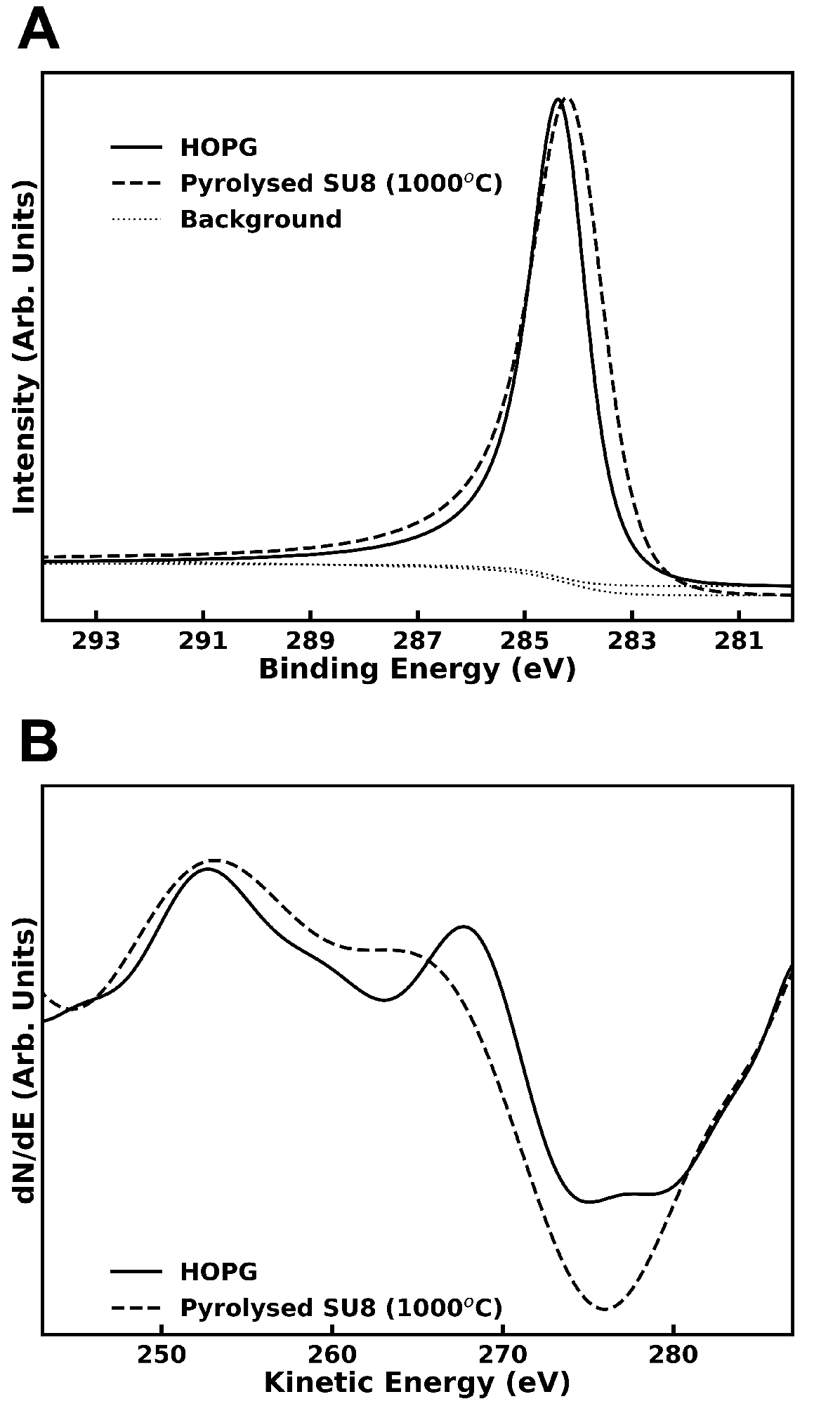}
    \caption{\textit{Comparison between HOPG and SU-8 pyrolysed at 1000~$\degree$C for both C 1s (Panel A) and first derivative C (KVV) Auger spectra (Panel B).}}    
    \label{fig:HOPG_compare}
\end{figure}

\noindent 
Highly ordered pyrolytic graphite (HOPG) was used as a comparison material \cite{schmieg_highly_1992} to confirm the quality of the glass-like carbon synthesised through SU-8 pyrolysis. HOPG was ex-situ cleaved and then heated to 300~$\degree$C in order to remove atmospheric contaminants on the surface. The HOPG sample was then allowed to cool in-situ and measured under the same experimental conditions as the SU-8 sample. Figure \ref{fig:HOPG_compare}A shows a comparison of the core level spectra for HOPG (solid line) and pyrolysed SU-8 (1000~$\degree$C, dashed line).  The two spectra have similar line shapes, with the HOPG exhibiting a narrower peak width. This is likely to be due to the single carbon environment within the HOPG;, however both exhibit similar values for asymmetry ($\sim$0.99). Figure \ref{fig:HOPG_compare}B shows the carbon KVV Auger spectra's first derivative (D-parameter) \cite{morgan_comments_2021, lascovich_evaluation_1991}. There are similar D parameters of 22.4 and 22.8 for HOPG and pyrolysed SU-8 respectively. This provides further evidence of the conversion from SU-8 to a material with a high concentration of ordered sp$^2$ carbon indicative of glass-like carbon.
\\\\
\noindent 
Both the peak position and peak asymmetry of the C-C peak in the C 1s spectra show a clear transition in the structure of the material at temperatures above 500~$\degree$C suggests that the development of a material that can be described as glass-like carbon begins at this temperature. The peak position and peak asymmetry plateau at temperatures above 750~$\degree$C, demonstrating these techniques to be suitable methods for determining the quality of pyrolysis of SU-8 into glass-like carbon.

\begin{figure} 
    \centering
    \includegraphics[scale=0.4]{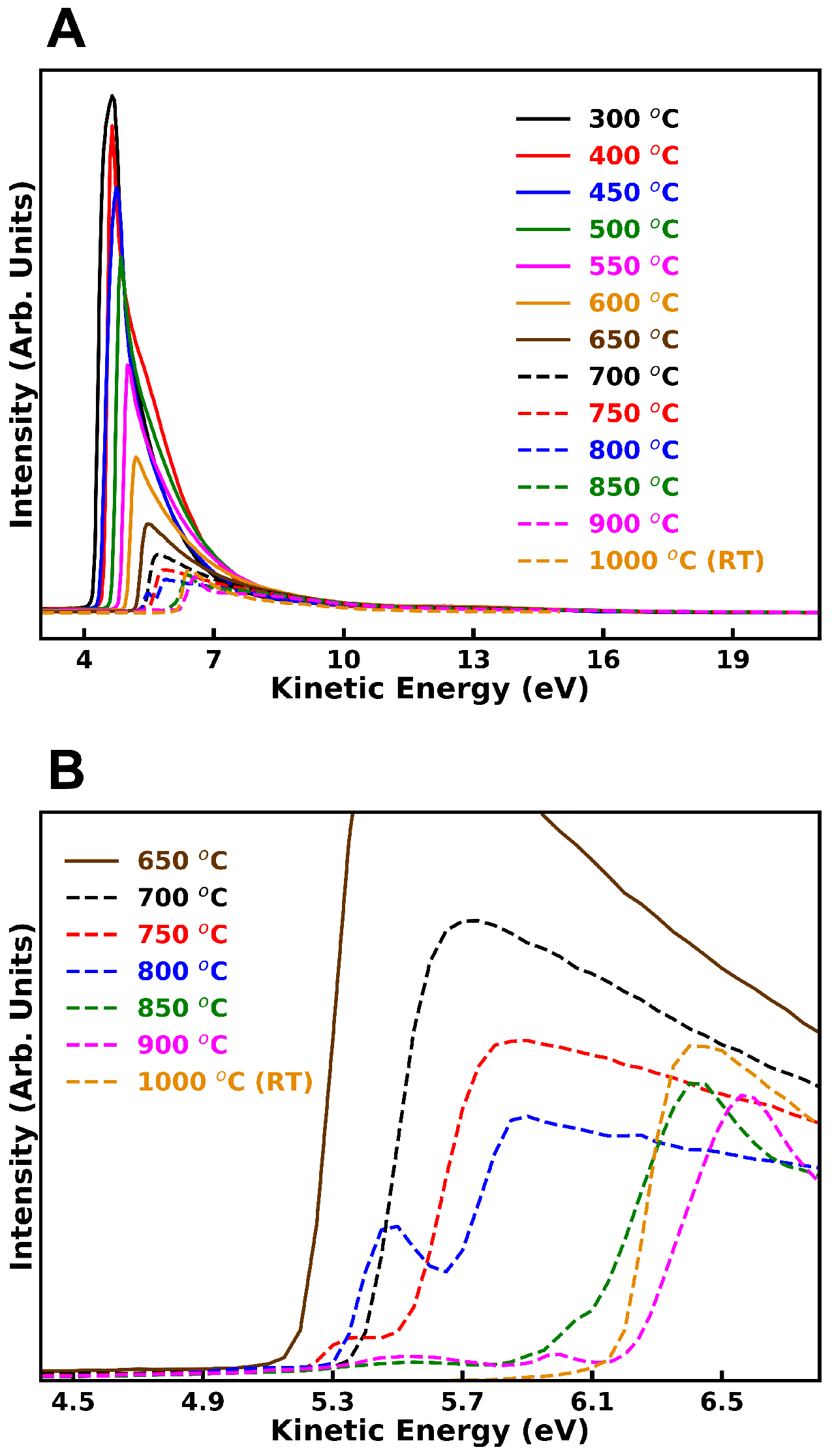}
    \caption{\textit{A) Full survey UPS data for SU-8 showing transition to glass-like carbon B) Zoomed in view of the survey spectra clearly showing the temperature region in which the transition occurs.}}    
    \label{ups}
\end{figure}

\noindent  
Figure \ref{ups}A and B show Ultraviolet Photoelectron Spectroscopy (UPS) spectra recorded at a range of temperatures between 300~$\degree$C and 1000~$\degree$C. The room temperature spectra (not shown) are dominated by surface contaminants that affect the more surface sensitive UPS measurements more than XPS. Contaminants are removed at 300~$\degree$C as suggested by the XPS core level data. The spectra shown in Figure \ref{ups}A are dominated by the emission of secondary electrons at low kinetic energy and their onset energy indicates a change of surface potential due to a changing work function (metal) or electron affinity (semiconductor). There is an overall increase in this secondary electron cut-off (SECO) energy with temperature and between 750~$\degree$C and 1000~$\degree$C there are changes in the pre-edge features that are revealed in the magnified SECO region in Figure \ref{ups}B. An additional peak appears at 750~$\degree$C that persists upto 950~$\degree$C before disappearing at 1000~$\degree$C. Additionally, a third peak appears around $\sim$6eV at 850~$\degree$C, developing into a full peak at 950~$\degree$C before finally disappearing at 1000~$\degree$C. These changes suggest changes in the surface structure and composition in this high temperature region that lead to changes in the surface potential. Multiple peaks in the SECO are often interpreted as the co-existence of different structures within the sampling region which, in this case is several mm$^2$ \cite{jang_su-8_2022, sharma2018}. 

\begin{figure} 
    \centering
   \includegraphics[scale=0.4]{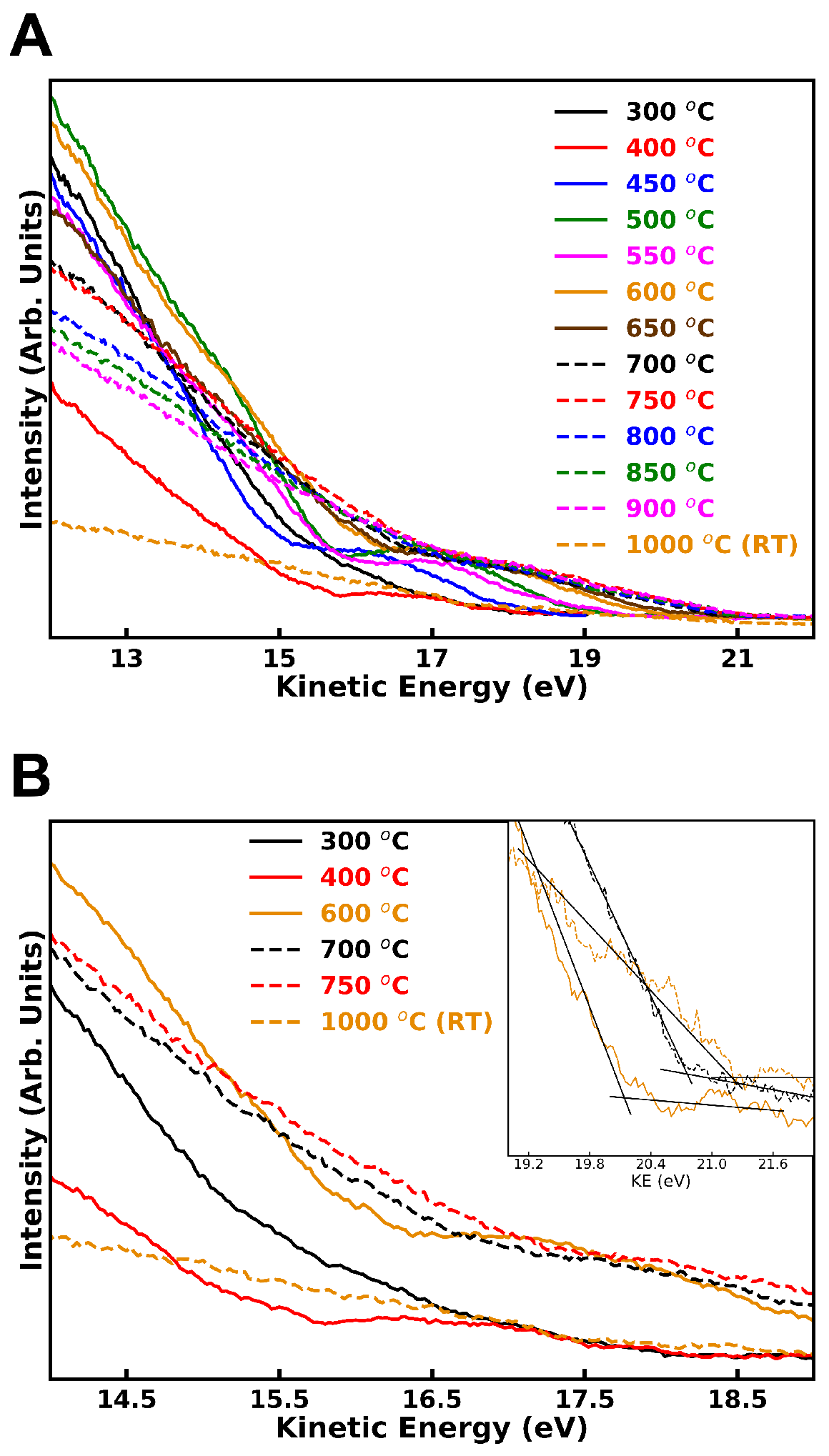}
    \caption{\textit{A) Valence band maximum of SU-8 taken at various temperatures indicating an insulating to conducting behaviour. B) Detailed view of the VBM showing the temperature range in which the transition occurs. The inset shows zoomed in curves at 600, 700 and 1000~$\degree$C showing the emergence of photoelectrons with binding energy close to zero.}}    
    \label{vbm}
\end{figure}

Figure \ref{vbm}A shows in more detail the valence band maximum of SU-8 as a function of temperature with spectra at selected temperatures shown in Figure \ref{vbm}B with a focus on photoelectrons emitted from near the Fermi Energy at 21.2 eV. The spectra were recorded with a small negative sample bias to enable measurement of the SECO, this value has been subtracted from the UPS spectra shown in Figures \ref{ups} and \ref{vbm}. As the temperature increases, a broad peak appears at 17 eV, that is most prominent at 400 ~$\degree$C and disappears in the temperature range of 700 - 750~$\degree$C. This coincides with the disappearance of O2 beyond 300~$\degree$C (Figure \ref{figo1s}) and the saturation of C-C peak position and is thus likely to be due to surface oxygen species (Figure \ref{c1s}). As the temperature approaches 1000~$\degree$C, the VBM shows increasing emission at the Fermi energy (zero binding energy) at 21.2 eV (inset Figure \ref{vbm}B) that indicates the full transformation of insulating SU-8 to fully conducting glass-like carbon reported in the literature \cite{sharma2018}.  

\section{Conclusions}
\label{sec:conclusions}
\noindent
X-ray photoelectron spectroscopy and ultraviolet photoelectron spectroscopy has been applied while heating in ultra-high vacuum conditions to investigate the structural and elemental compositional changes in SU-8 photoresist as it pyrolyses to glass-like carbon in real time. Elemental analysis through survey scans and analysis of detailed O 1s scans show a clear reduction in structural oxygen between 300 - 600~$\degree$C, with gaseous oxygen desorbing from the structure slowly due to the  reduced porosity of glass-like carbon. A clear transition in the pyrolysis mechanism at temperatures above 500~$\degree$C is manifested through increasing electrical conductivity due to the development of a conducting sp$^2$ carbon network. This is confirmed by the increase in C 1s peak asymmetry and the similarity of the asymmetry and the Auger parameter of the glass-like carbon with HOPG. The UPS VBM measurements further confirm the transition to a conducting phase as demonstrated by photoelectron emission at the Fermi energy at the highest temperature and the SECO reveals structural changes that lead to surface phases with different ionisation energies at the higher temperatures.

\section*{Acknowledgements}
\noindent
We gratefully acknowledge the Engineering and Physical Sciences Research Council (EPSRC) for funding this project under the "Making a miniature Sun" grant (EPSRC Reference EP/V048295/1). X-ray photoelectron (XPS) data were acquired at the Aber XPS facility, as a part of the EPSRC National Facility for XPS (“HarwellXPS”, EP/Y023587/1, EP/Y023609/1, EP/Y023536/1, EP/Y023552/1 and EP/Y023544/1)' 

\section*{CRediT authorship contribution statement}
Simon Astley: Writing – review \& editing, Methodology, Investigation, Formal analysis, Data curation. Jaspa Stritt: Writing – review \& editing, Writing – original draft, Visualization, Methodology, Investigation, Formal analysis, Data curation. Soumen Mandal: Writing – review \& editing, Writing – original draft, Supervision, Visualization, Methodology, Investigation, conceptualization.  Jerome A. Cuenca: Writing – review \& editing, Supervision, Methodology, Investigation, Formal analysis, conceptualization. D. Andrew Evans: Writing – review \& editing, Supervision, Methodology, Investigation, Formal analysis. Oliver A. Williams: Writing – review \& editing, Supervision, Resources, Project administration, Methodology, Funding acquisition, Conceptualization.

\section*{Data availability}
Information on the data underpinning this publication, including access details, can be found in the Cardiff University Research Data Repository at https://doi.org/10.17035/cardiff.29931761.v1


 \bibliographystyle{elsarticle-num} 
 \bibliography{ref1}





\end{document}